\documentclass[twocolumn]{aastex631}

\newcommand\lsim{\mathrel{\rlap{\lower4pt\hbox{\hskip1pt$\sim$}}
\raise1pt\hbox{$<$}}}
\usepackage{xcolor}

\shorttitle{Turning Points}
\shortauthors{Rose \& MacLeod 2023}

\graphicspath{{./}{figures/}}

\begin{document}

\title{
Collisional Shaping of Nuclear Star Cluster Density Profiles} 

\correspondingauthor{Sanaea C. Rose}
\email{sanaea.rose@northwestern.edu}

\author[0000-0003-0984-4456]{Sanaea C. Rose}
\affiliation{Center for Interdisciplinary Exploration and Research in Astrophysics (CIERA), Northwestern University, 1800 Sherman Ave, Evanston, IL 60201, USA}
\affiliation{Department of Physics and Astronomy, University of California, Los Angeles, CA 90095, USA}
\affiliation{Mani L. Bhaumik Institute for Theoretical Physics,
University of California, Los Angeles,
CA 90095, USA}

\author[0000-0002-1417-8024]{Morgan MacLeod}
\affiliation{Center for Astrophysics, Harvard \& Smithsonian, 60 Garden Street, MS-16, Cambridge, MA, 02138, USA}

\begin{abstract}
A supermassive black hole (SMBH) surrounded by a dense, nuclear star cluster resides at the center of many galaxies. In this dense environment, high-velocity collisions frequently occur between stars. About $10 \%$ of the stars within the Milky Way's nuclear star cluster collide with other stars before evolving off the main-sequence. Collisions preferentially affect tightly-bound stars, which orbit most quickly and pass through regions of the highest stellar density.  Over time, collisions therefore shape the bulk properties of the nuclear star cluster. We examine the effect of collisions on the cluster's stellar density profile. We show that collisions produce a turning point in the density profile which can be determined analytically. Varying the initial density profile and collision model, we characterize the evolution of the stellar density profile over $10$~Gyr. We find that old, initially cuspy populations exhibit a break around $0.1$~pc in their density profile, while shallow density profiles retain their initial shape outside of $0.01$~pc. The initial density profile is always preserved outside of a few tenths of parsec irrespective of initial conditions. Lastly, we comment on the implications of collisions for the luminosity and color of stars in the collisionly-shaped inner cluster.
\end{abstract}


\section{Introduction}

Most galaxies harbor a supermassive black hole at their center \citep[e.g.][]{FerrareseFord05,KormendyHo13}. A dense cluster of stars and stellar remnants surrounds these SMBHs \citep[e.g.,][]{Morris93,schodel+03,Ghez+05,Ghez+08,Gillessen+09,Gillessen+17,Neumayer+20}. The proximity of the Milky Way's galactic nucleus (GN) presents a unique opportunity to study the populations of stars and compact objects surrounding a SMBH. The stellar density profile in particular can tell us about the dynamical history of the GN \citep[e.g.,][]{Baumgardt+06,LockmannBaumgardt08,Merritt2010,Bar-Or+13,Mastrobuono-Battisti+14}.

Within the GN, stars trace orbits dominated by the gravity of the SMBH \citep[e.g.,][]{Ghez+08,Genzel+10}. A power law of the form $\rho \propto r^{-\alpha}$ is often used to describe the stellar mass density as a function of distance $r$ from the SMBH. In this dense environment, stars also experience weak gravitational interactions with one another. These interactions allow the stars to exchange  energy over time. This process, called relaxation, redistributes the stellar orbits onto an equilibrium density profile. Theoretical models predict that an old, relaxed population should have a cuspy density profile with  slope $\alpha$ lying between $1$ and $1.75$, depending on what is assumed about the star formation history and relative abundances of stars and compact objects \citep[e.g.,][]{BahcallWolf76,AharonPerets16,LinialSari22}.

Contrary to expectations, however, the old stellar population in the GN, traced using bright evolved giants, does not appear to have a cuspy density profile within $\sim 0.1$~pc of the SMBH \citep[e.g.,][]{Genzel+96,BaileyDavies99,Buchholz+09,Do+09,Do+13a,Do+13b,Baumgardt+18,Habibi+19RG}. These observations have prompted several proposed mechanisms to preferentially destroy red giants at these radii \citep[e.g.,][]{Davies+98,Alexander99,BaileyDavies99,Dale+09,AmaroSeoane&Chen14,Zajacek+20,AmaroSeoane+20RGs}. It is therefore possible that the red giant density profile does not trace the distributions of the main-sequence stars or compact object populations. Recent observational campaigns suggest that the main-sequence stars lie on a cusp, albeit a shallower one with index $\alpha$ between $1.1$ and $1.4$ \citep[][]{Gallego-Cano+18,Schodel+18}. However, observations of the stellar cusp are challenging \citep[e.g.,][]{Schodel+20}, and there may be connections between the main-sequence stellar density profile and the products of their evolution \citep[e.g.,][]{Rose+23}. Specifically, the core-like profile of the bright evolved stars internal to $\sim 0.1$~pc may have a dynamical origin \citep[e.g.,][]{Merritt2010,Rose+23}. 

We explore the potential of one such dynamical process, direct collisions between stars, to shape the stellar density profile.
In the dense environment of a nuclear star cluster, direct collisions between objects become possible \citep[e.g.,][]{DaleDavies,Dale+09,2011AdAst2011E..13R,Mastrobuono-Battisti+21,Rose+20,Rose+22,Rose+23}. These collisions have been studied in a variety of contexts in the literature, including AGN variability \citep[e.g.,][]{Murphy+91,Torricelli-C+00,FreitagBenz02}, electromagnetic and gravitational wave signals \citep[e.g.,][]{DaleDavies,Balberg+13,AmaroSeoane23}, and the presence of young-seeming, bright stars \citep[e.g.,][]{Sills+97,Sills+01,Lombardi+02,Rose+23}. Several theoretical and computational studies have shown that destructive collisions can deplete the supply of stars near the SMBH \citep[e.g.,][]{DuncanShapiro83,Murphy+91,David+87a,David+87b,Rauch99,FreitagBenz02,Rose+23}. The frequency and outcomes of direct collisions depends on distance from the SMBH \citep[e.g.,][]{Lai+93,Rauch99,2011AdAst2011E..13R,2021arXiv210514026H,Rose+23}. This process may therefore have distinct effects on the stellar density profile \citep[e.g.,][]{Rauch99,FreitagBenz02}.


We leverage a toy model developed by \citet{Rose+22,Rose+23} to examine the effects of stellar collisions on the density profile of a GN using the Milky Way's as an example \citep[e.g.,][]{Ghez+05}. We vary the initial density profile and the collision model to build a comprehensive picture of possible evolutions of the density profile, the circumstances under which a break in the profile arises, and regions of the nuclear star cluster in which the original profile is preserved. 

This paper is organized as follows.
In Section~\ref{sec:method}, we provide an overview of our model and the key dynamical processes considered. Section~\ref{sec:characteristicradii} provides an analytic framework to understand collisional shaping of the density profile and to aid in the interpretation of our simulated results. Section~\ref{sec:density} presents and discusses the evolution of the density profile for different simulations, while Section~\ref{sec:luminosity} discusses implications for the luminosity profile. We conclude in Section~\ref{sec:conclusions}.







\section{A Nuclear Star Cluster Model} \label{sec:method}

This section describes our semi-analytic approach to modeling the stellar surroundings of a SMBH. Our model adopts a simplified description of the stellar density profile and dynamics around the SMBH, but includes the effects of star--star collisions. We adopt conditions representative of the Milky Way's GN, but these models can be easily adapted to other GN.

\subsection{Nuclear Star Cluster Properties}


In our model nuclear star cluster, the stellar mass density is described by a power law:
\begin{eqnarray} \label{eq:density}
    \rho(r_\bullet) = \rho_0 \left( \frac{r_\bullet}{r_0}\right)^{-\alpha} \ , 
\end{eqnarray}
where $\alpha$ is the slope and $r_\bullet$ denotes distance from the SMBH. The density profile within the sphere of influence of the SMBH is normalized using $\rho_0 = 1.35 \times 10^6 \, M_\odot/{\rm pc}^3$ at $r_0 = 0.25 \, {\rm pc}$, based on observations of this region \citep{Genzel+10}. In our simulations, we test three of values of $\alpha$, $1.25$, $1.5$, and $1.75$, to encapsulate the range of theoretical predictions and observed density profiles.

The velocity dispersion within the cluster is also a function of distance from the SMBH:
\begin{eqnarray}\label{eq:sigma}
    \sigma(r_\bullet) = \sqrt{ \frac{GM_{\bullet}}{r_\bullet(1+\alpha)}},
\end{eqnarray}
where $\alpha$ is the slope of the density profile and $M_{\bullet}$ is the mass of the SMBH \citep{Alexander99,AlexanderPfuhl14}. We set $M_\bullet$ equal to $4 \times 10^6$~M$_\odot$, the mass of the Milky Way's SMBH \citep[e.g.,][]{Ghez+03}. For a uniform mass cluster of $1$~M$_\odot$ stars, the number density $n$ is simply $\frac {\rho(r_\bullet)}{1 \, M_\odot}$. Together, the density and velocity dispersion in the nuclear star cluster determine the frequency of interactions between stars and set key timescales for various dynamical processes.


\subsection{Overview of Dynamical Processes}

\subsubsection{Collision Rate}
In dense star clusters, direct collisions between objects become possible \citep[e.g.,][]{DaleDavies,Dale+09,2011AdAst2011E..13R,Mastrobuono-Battisti+21,Rose+22,Rose+23}. The timescale for a direct collision can be estimated as, $t_\mathrm{coll}^{-1} = n \sigma A$, where $A$ is the cross-section of interaction, $n$ is the number density, and $\sigma$ is the velocity dispersion, here a measure of the relative velocity between objects. For a direct collision, the cross-section of interaction $A$ is the physical cross-section enhanced by gravitational focusing. The eccentricity of a star's orbit about the SMBH, $e_\bullet$, can also affect the collision timescale; more eccentric orbits have shorter collision timescales compared to circular orbits with the same semimajor axes \citep[e.g.,][]{Rose+20}. Including the eccentricity dependence, the collision timescale can be written as:
\begin{eqnarray} \label{eq:t_coll_main_ecc}
     t_{\rm coll}^{-1} &=& \pi n(a_\bullet) \sigma(a_\bullet) \nonumber \\ &\times& \left(f_1(e_\bullet)r_c^2 + f_2(e_\bullet)r_c \frac{2G(M_\odot+M)}{\sigma(a_\bullet)^2}\right)\ .
\end{eqnarray}
where $f_1(e_\bullet)$ and $f_2(e_\bullet)$ are given by \citet{Rose+20} equations 20 and 21, $G$ is the gravitational constant, $a_\bullet$ is the semimajor axis of the star's orbit, and $r_c$ is the sum of the radii of the colliding stars. We plot the collision timescale in red in Figure~\ref{fig:timescales} for the range of density profiles considered in this study. $\alpha = 1.75$ is shown in the solid line, while $\alpha = 1.25$ is shown in the dashed line. This timescale assumes a uniform population of solar mass stars. Therefore, $r_c = 2R_\odot$. 

\subsubsection{Two-Body Relaxation}

Even more frequent than direct physical collisions are weak gravitational interactions between passing stars. These interactions cause the orbital parameters to change slowly over time in a diffusion process. Over the so-called relaxation timescale, the star's orbital energy and angular momentum change by order of themselves. Like the collision timescale, the relaxation timescale $t_\mathrm{rlx}$ depends on properties of the cluster such as the density and velocity dispersion, both functions of $r_\bullet$. The timescale can be expressed as:
\begin{eqnarray} \label{eq:t_rlx}
t_{\rm rlx} = 0.34 \frac{\sigma^3}{G^2 \rho \langle M_\ast \rangle \ln \Lambda_{\rm rlx}} \ ,
\end{eqnarray}
where $\langle M_\ast \rangle$ is the average mass of the objects in the cluster, here taken to be $1$~M$_\odot$, and $\ln \Lambda_{\rm rlx}$ is the coulomb logarithm \citep[e.g.,][]{BinneyTremaine,Merritt2013}. 
We plot this timescale in blue in Figure~\ref{fig:timescales} for a range of density profiles, spanning $\alpha = 1.25$ (dashed) to $1.75$ (solid). The timescale is less than or comparable to the duration of our simulations, $10$~Gyr, shown in grey in Figure~\ref{fig:timescales}. Additionally, while outside of $0.1$~pc, the collision timescale is long compared to the total simulation time, relaxation processes can change the orbits of the stars and move them into regimes where collisions become likely. The reverse is also true: stars that begin in regions where collisions are common can migrate further from the SMBH. It is therefore important to account for this diffusion process in our semi-analytic models.

\subsection{Semianalytic Model}
We use a toy model developed by \citet{Rose+22,Rose+23} to simulate the effects of collisions on the star cluster. This model follows a subset of $1000$ stars of varying masses, drawn from a Kroupa initial mass function (IMF), embedded in a fixed cluster of $1$~M$_\odot$ stars. The index $\alpha$ of the surrounding star cluster's density profile is treated as a free parameter with a default value of $1.75$, the expectation for an old, dynamically relaxed population \citep[e.g.,][]{BahcallWolf76}. The orbital eccentricities of the stars in our sample are drawn from a thermal distribution, while their semimajor axes are drawn from a uniform distribution in log distance from the SMBH.

The code takes a statistical approach to collisions. It computes to probability of a collision occurring over a timestep $\Delta t$, taken to be $10^6$~yr, as $\Delta t/t_\mathrm{coll}$. We then draw a random number between $0$ and $1$. If the number is less than or equal to the collision probability, we treat the star as having collided, and update its properties given the models described in Section \ref{sec:collisionmodel}. This prescription repeats until the code has reached the desired simulation time, $10$~Gyr, or the star has reached the end of its main-sequence lifetime, whichever occurs first. We also simulate the effects of relaxation in our code. Over each timestep, we apply a small instantaneous velocity kick to the star, from which we calculate the new, slightly altered orbital parameters. The velocity kick is drawn from a Guassian distribution with a standard deviation that depends on the ratio of the orbital period to the relaxation timescale for the star in question \citep[e.g.,][see the latter for the full set of equations]{Bradnick+17,Lu+19,Rose+22,Rose+23,Naoz+22}. This prescription allows us to account for diffusion in the orbital parameters over time from interactions with the surrounding stars.

\subsection{Treatment of Collision Outcomes} \label{sec:collisionmodel}

If a collision occurs, we must adjust the mass of the star in our sample accordingly. The collision outcome depends in part on the speed of the impact. The velocity dispersion in the nuclear star cluster exceeds $100$~$\mathrm{km}/\mathrm{s}$ within about $0.1$~pc of the SMBH. Heuristically, high velocity collisions should result in mass loss from the colliding stars. In extreme cases, when the relative velocity is larger than the escape speed from the star, one might expect the collision to fully unbind the star.

\citet{Rose+23} find that the collision outcomes can generally be understood in three regimes. Near the SMBH, the velocity dispersion exceeds the escape velocity from the stars, leading to destructive collisions with high mass loss. Between about $0.01$-$0.1$~pc, collisions are common and can lead to mergers. These merger products, more massive than their progenitor stars, evolve off the main-sequence over a shorter lifetime. Outside of $0.1$~pc, collisions are less frequent, but always lead to mergers with little mass loss. 

In more detail, collision outcomes depend on the gas dynamics of the collisions themselves and are dependent on several other conditions beyond the impact velocity, such as the impact parameter and the internal structure of the stars \citep[e.g.,][]{Lai+93,FreitagBenz02,2011AdAst2011E..13R}. Previous studies have leveraged hydrodynamic simulations to understand collision outcomes at high velocities for different impact parameters \citep[e.g.,][]{Lai+93,Rauch99,FreitagBenz02}. \citet{Lai+93} and \citet{Rauch99}, in particular, provide fitting formulae based on their results. These formulae can be easily implemented in a code such as ours to estimate the mass lost from the stars and whether or not a merger occurred \citep{Rose+23}.

In this study, we compare three different recipes for determining the collision outcome.
We begin by considering a limiting case in which collisions are fully destructive. In this prescription, the occurrence of a single collision terminates the code. Once a star in our sample experiences a collision, its mass is set to zero and it is removed from the population. While unphysical, this treatment of collisions allows us to build an intuition for the underlying physics of our models. We then proceed to include a more complicated treatment of the collision outcomes. These simulations utilize either fitting formulae from \citet{Rauch99} or \citet{Lai+93}, discussed in more detail in \citet{Rose+23}. Henceforth, we refer to simulations that use fitting formulae from these studies as ``Rauch99'' and ``Lai+93'', respectively. These fitting formulae allow us to calculate the mass loss from a given collision given the mass ratio of the colliding stars, the impact parameter, which is drawn statistically, and the relative velocity. Our simulations always assume that the relative velocity is equal to the velocity dispersion, a function of distance from the SMBH given by Eq.~(\ref{eq:sigma}). As described in \citet{Rose+23}, our Rauch99 simulations favor mergers with little mass loss, while collisions lead to higher mass loss in the Lai+93 simulations and mergers are less likely. Together, these prescriptions are selected to span the range of possible collision outcomes.




\section{Characteristic Radii} \label{sec:characteristicradii}

\begin{figure}
	\includegraphics[width=0.99\columnwidth]{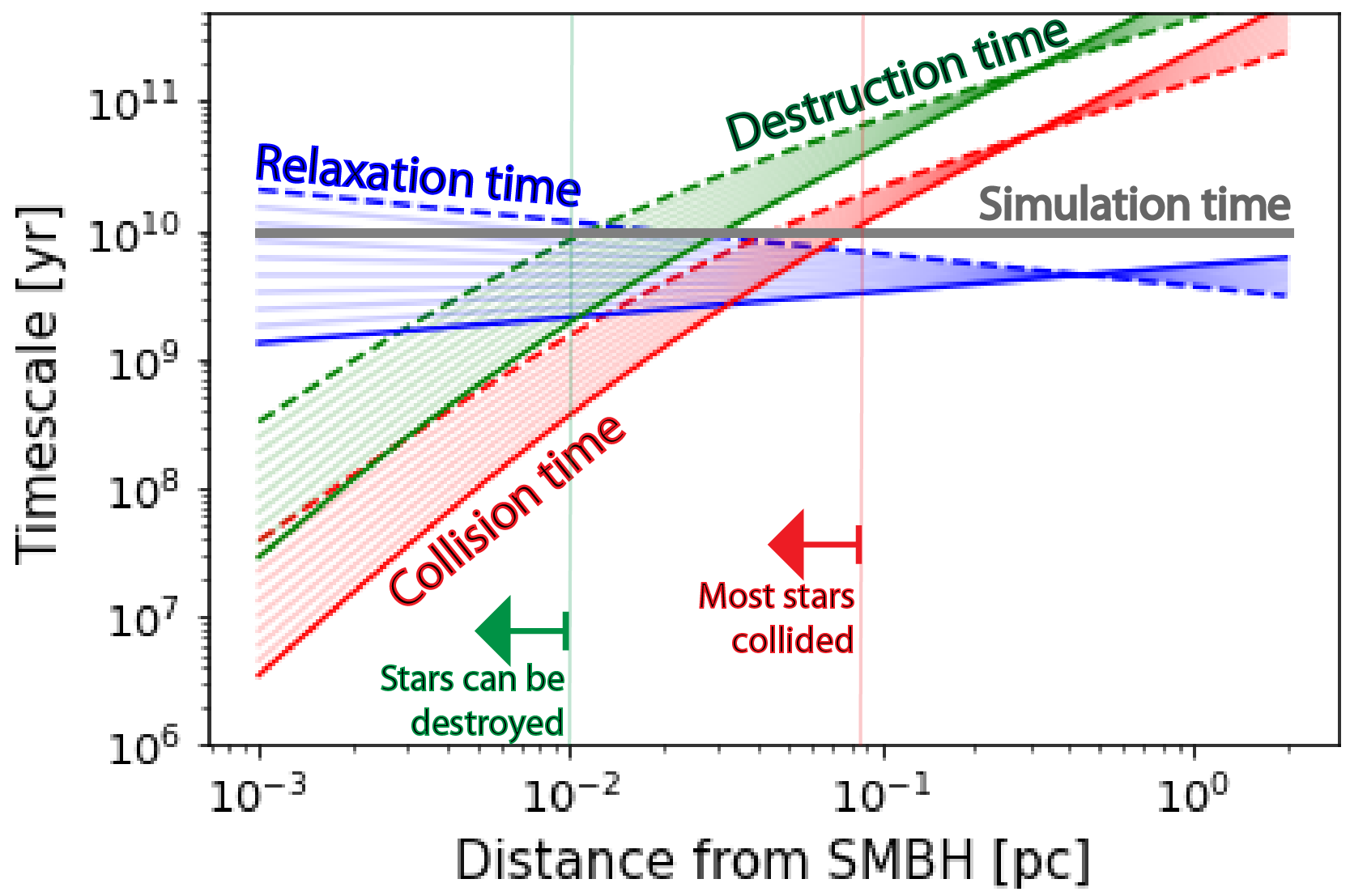}
	\caption{Assuming a uniform population of $1$~M$_\odot$ stars, we plot relevant timescales for a range of stellar density profiles, $\alpha = 1.25$ (solid line) to $\alpha = 1.75$ (dashed line), in the nuclear star cluster . The collision and relaxation timescales are in red and green, respectively. We also include a destruction timescale, approximately the time needed for the two stellar cores to collide, or the collision timescale (Eq.~\ref{eq:t_coll_main_ecc}) calculated for $r_c = 2 \times 0.33 R_\odot$. Within about $0.01$~pc of the SMBH, the kinetic energy is sufficiently high that a collision with small impact parameter can unbind the stars. To guide the eye, the grey line shows the total simulation time of $10$~Gyr.
 }
     \label{fig:timescales}
\end{figure}

As mentioned above, the relevance of various physical processes becomes clear when one compares their associated timescales to the duration of our simulations. We show two key timescales, relaxation in green and collision in red, in Figure~\ref{fig:timescales} along with the simulation time $10$~Gyr (grey). We can define characteristic radii in the nuclear star cluster by equating various timescales.

\subsection{The Collision Radius}

Collisions play a crucial role in shaping the stellar demographics where the collision timescale, $t_\mathrm{coll}$, is less than the age of the population. Setting $t_\mathrm{coll}$ equal to $t_\mathrm{age}$ gives a critical radius, $r_\mathrm{coll}$,  within which the vast majority of the stars have collided. Outside of $r_\mathrm{coll}$, a fraction of the stellar population will still experience collisions. This fraction can be estimated using $t_\mathrm{coll}/t_\mathrm{age}$ \citep[see also figure 1 in][]{Rose+23}. In addition to population age, $r_\mathrm{coll}$ also depends on the steepness of the density profile. For example, the age of an old $10$~Gyr population and collision timescale intersect closer to $0.04$~pc for $\alpha = 1.25$, compared to $\sim 0.1$~pc for $\alpha = 1.75$.

This analysis informs where in the nuclear star cluster we expect collisions to be an important process in shaping the cluster properties. Because collisions modify or destroy stars at high enough velocity, we predict that $r_\mathrm{coll}$ will correspond to a break in the stellar density profile. Within $r_\mathrm{coll}$, collisions are an important process in determining the stellar density profile. Outside of this critical radius, collisions are rare, and the density profile is not shaped by collisions. Over time, as the age of the population increases, the inflection point will move further from the SMBH. For an old $10$~Gyr population, with properties similar to the Milky Way GN, $r_\mathrm{coll}$ occurs at about $0.1$~pc, shown in Figure~\ref{fig:timescales}.

\subsection{The Destruction Radius}

We can perform a similar analysis to determine where in the cluster collisions can effectively deplete the entire supply of stars. Within about $0.01$~pc of the SMBH, the velocity dispersion, given by Eq.~\ref{eq:sigma}, exceeds the escape velocity from a Sun-like star. In this region, collisions have the potential to destroy the stars. About two thirds of the Sun's mass is concentrated in the inner third of its radius \citep[e.g.,][]{Christensen-Dalsgaard+96}. A collision will result in high mass loss when the impact parameter is small enough that the dense cores interact \citep[e.g.,][]{Lai+93,Rauch99,Rose+23}. We define a characteristic timescale over which the stars will be destroyed by setting the impact parameter $r_c$ equal to $0.33$~R$_\odot$. Figure~\ref{fig:timescales} shows this timescale in green. This timescale is consistent with \citet{Rose+23}, who find the time needed to deplete the stellar population within $0.01$~pc to be about a Gyr. Similar to $r_\mathrm{coll}$, we define $r_\mathrm{dest}$ as the radius at which the destruction timescale equals the population age. We stress that this definition is only valid in regions where the collision velocity is high enough to destroy the stars.

\begin{figure*}
	\includegraphics[width=0.99\textwidth]{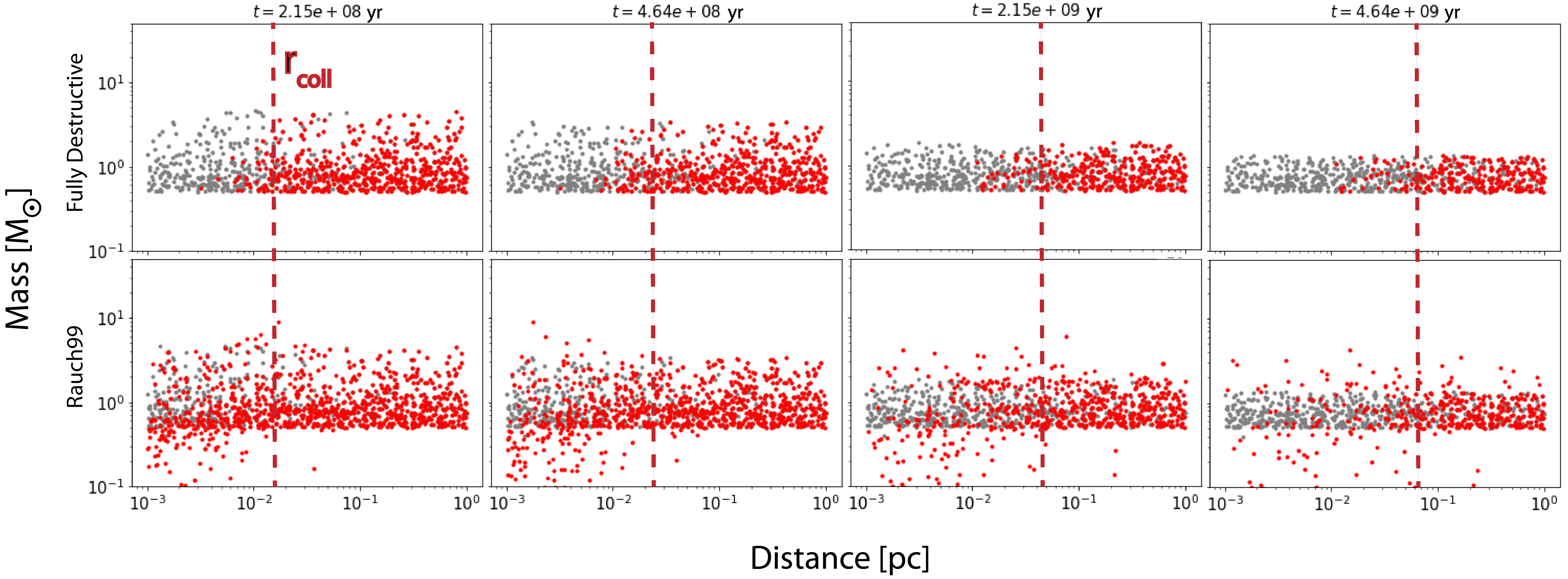}
	\caption{The stellar population at four different times over the course of two simulations. As indicated by the column labels, time increases left to right. The grey points show the masses of the stars at the given time as determined by main-sequence evolution alone. The red dots show the simulated masses of the stars in our sample, which can also change due to collisions. The top row corresponds to fully destructive collisions, meaning a star is removed from the sample when it undergoes a single collision. The bottom row shows results that use the mass loss prescription from \citet{Rauch99}. Both simulations assume a Bahcall-Wolf ($\alpha = 1.75$) profile for the surrounding stars. We mark $r_\mathrm{coll}$ with a red dashed vertical line, where $r_\mathrm{coll}$ is calculated from Eq.~\ref{eq:scale2}.
 }
     \label{fig:snaps}
\end{figure*}

\subsection{Generalizing to Other Galactic Nuclei}

The break radius for any GN can be found by equating the population age and collision timescale: $t_\mathrm{age} = (n\sigma A)^{-1}$. Similar to the Milky Way's GN, the SMBH of mass $M_\bullet$ dominates the gravitational potential within the sphere of influence, and the relative velocity $\sigma$ can be calculated using Eq.~\ref{eq:sigma}. The initial stellar density profile of the cluster must be calibrated to the mass of the central SMBH. Using the $M$-$\sigma$ relation, the stellar density profile for a GN with a SMBH of arbitrary mass can be written as a power law:
\begin{eqnarray}
    \rho(r_\bullet) = \frac{3-\alpha}{2\pi} \frac{M_\bullet}{r_\bullet^3} \left( \frac{G(M_0M_\bullet)^{1/2}} {\sigma_0^2 r_\bullet} \right)^{-3+\alpha} \ , \label{eq:density_allM}
\end{eqnarray}
where $M_0=1.3 \times 10^8 \, M_\odot$, $\sigma_0 = 200 \mathrm{km} \mathrm{s}^{-1}$, and $r_\bullet$ is the distance from the SMBH \citep{Tremaine+02}. 

We can derive a scaling relation using $r_\mathrm{coll} = 0.09$~pc for the Milky Way's GN at $10$~Gyr for a Bahcall-Wolf profile ($\alpha = 1.75$). For a GN with arbitrary stellar density profile slope $\alpha$, the break radius can be calculated using the following:
\begin{eqnarray}
    r_\mathrm{coll} &=& \left[ 1.49 \times 10^{-6} \left(6.04\right)^\alpha \frac{(3-\alpha)^2}{1+\alpha}\right]^{\frac{1}{1+2\alpha}} \nonumber  \\ &\times& \left( \frac{t_\mathrm{age}} {10^{10}~\mathrm{yr}} \right)^{\frac{2}{1+2\alpha}} \left( \frac{M_\bullet}{4 \times 10^6 M_\odot}\right)^{\frac{\alpha}{1+2\alpha}} \mathrm{pc}. \label{eq:scale1}
\end{eqnarray}
For values of $\alpha$ between $1$ and $2$, the first term in brackets changes by a factor $\sim 5.4$. The following equation can serve as an approximation in this range of $\alpha$:
\begin{eqnarray}
    r_\mathrm{coll} &\sim& 0.07 \left( \frac{t_\mathrm{age}} {10^{10}~\mathrm{yr}} \right)^{\frac{2}{1+2\alpha}} \left( \frac{M_\bullet}{M_\odot}\right)^{\frac{\alpha}{1+2\alpha}} \mathrm{pc}. \label{eq:scale_simple}
\end{eqnarray}
If the GN also has a Bahcall-Wolf profile, i.e., $\alpha = 1.75$, the full relation can instead be simplified to:
\begin{eqnarray}
    r_\mathrm{coll} = 0.09 \left( \frac{t_\mathrm{age}} {10^{10}~\mathrm{yr}} \right)^{0.44} \left(\frac {M_\bullet} {4 \times 10^6~\mathrm{M}_\odot} \right)^{0.39} \mathrm{pc} \ , \label{eq:scale2}
\end{eqnarray}
We use this last relation to calculate $r_\mathrm{coll}$ in our fiducial models, setting $M_\bullet = 4 \times 10^6$~M$_\odot$.


\section{Numerical Results} \label{sec:results}

Here, we present results from several simulations with various initial stellar density profiles and collision outcome prescriptions, reviewed in Section~\ref{sec:collisionmodel}. We discuss these results in the context of the stellar density and luminosity profiles, and we compare the simulations to our predictions from Section~\ref{sec:characteristicradii}.

\subsection{Collisional Shaping of Stars}\label{sec:stars}

Figure \ref{fig:snaps} shows the time evolution of the model population of ``tracer" stars. We show four snapshots for two different simulations, both of which adopt $\alpha = 1.75$ for the density profile. The top row assumes that collisions are fully destructive, while the bottom row uses a Rauch99 prescription for collision outcomes. The grey points in the plots show the stellar masses at a given time due to stellar evolution alone. As can be seen in both rows of the figure, the vertical extent of the grey points decreases with time because the time elapsed has exceeded the main-sequence lifetime of stars above a mass threshold. 

The red points show the stellar masses from our simulations, which account for the effects of collisions. We mark the of location $r_\mathrm{coll}$ with a dashed red vertical line in each snapshot, the distance within which the majority of the stars have collided. Calculated from Eq.~\ref{eq:scale2}, $r_\mathrm{coll}$ sweeps out over time.

Depending on the collision prescription and the impact velocity, collisions can either add or remove mass from the impacting stars \citep[as described in detail by][]{Rose+23}.  In the top row, collisions carve out a star-less region over time. The bottom row, on the other hand, incorporates a more complete and complex treatment of collision outcomes. Thus, the region inside $r_{\rm coll}$ contains both collisionally-merged,  more massive stars and collisionally-stripped, less massive stars as compared to their progenitors.

\subsection{Cluster Density Profile} \label{sec:density}

\begin{figure*}
	\includegraphics[width=0.99\textwidth]{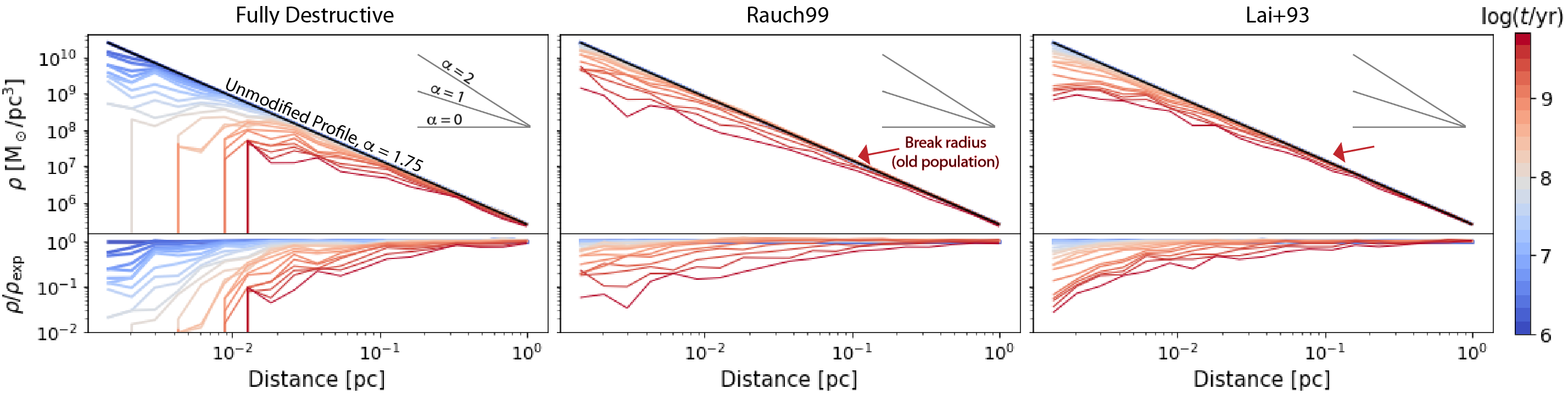}
	\caption{The evolution of the stellar density profile for different simulations. The title of each column states the collision outcome prescription used in the simulation shown below. The black lines represent the initial, unmodified density profile, with $\alpha = 1.75$. The colored lines show the profile at a given time, indicated by the colorbar on the right. The grey lines in the upper left corner of each plot show other examples of density profile slopes to guide the eye. The left column presents results for fully destructive collisions. These density profiles exhibit clear turning points coinciding with where the time elapsed, or population age, equals the collision timescale (see Figure~\ref{fig:timescales}). In the other rows, a more complex treatment of collision outcomes obscures the inflection point in the density profiles. However, a similar trend is present: the inflection point moves further from the SMBH with time, corresponding roughly to where the population age equals the collision timescale. The red arrows draw attention to this approximate distance from the SMBH for an old ($\sim 10$ Gyr) population.
 }
     \label{fig:densityprofiles}
\end{figure*}

\begin{figure*}
	\includegraphics[width=0.99\textwidth]{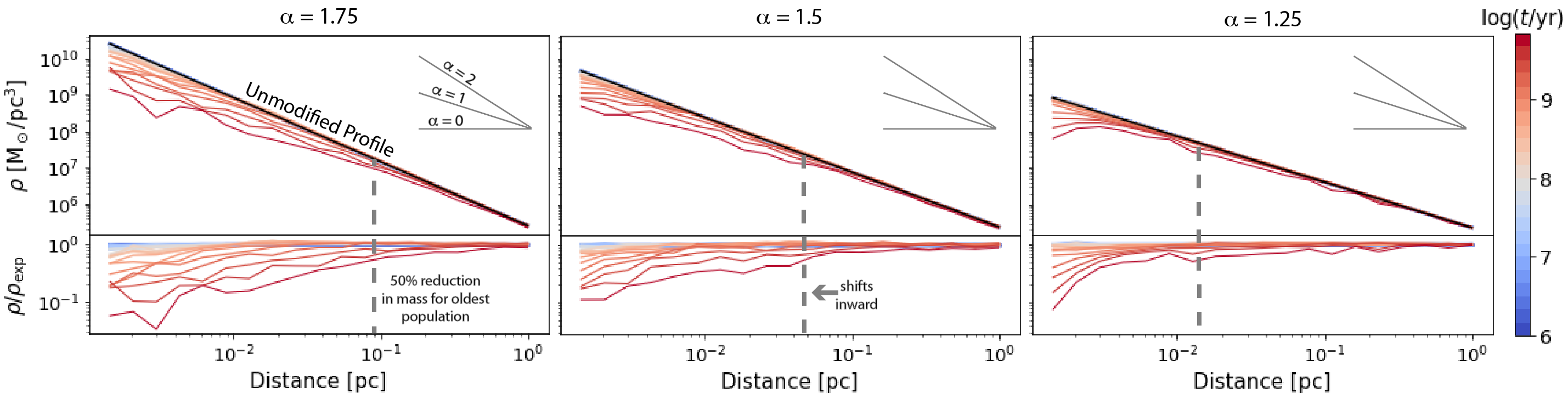}
	\caption{The stellar density profile with varying initial density profile of the stars. The title of each column indicates the value of $\alpha$ used for the stellar density profile. Each plot has the same axes to facilitate a comparison between them. The form of the evolving density profiles suggest that the initial profile is always preserved outside of $0.2$~pc, where collisions are more rare. A steeper the initial profile, however, also results in a more pronounced break in the density profile at a further distance from the SMBH. To highlight this trend, we have added a dashed grey line marking where the mass density is halved due to collisions for the oldest population shown in the figure (darkest red curve, $7$~Gyr).
 }
     \label{fig:differentalpha}
\end{figure*}

We consider the evolution of the stellar density profile over time due to collisions, which can work to either destroy or merge stars. To generate a density profile, we slice up each snapshot, including those depicted in Figure~\ref{fig:snaps}, in distance. In each annulus with width $\delta r$ that lies distance $r$ from the SMBH, we divide the total mass in the red points by the total mass in the grey points, which tells us the fractional change in mass at that distance at a particular time. We then convolve those fractions with the original density profile of the stars to obtain a collisionally-modified profile.

We juxtapose the evolving density profiles of three simulations in Figure~\ref{fig:densityprofiles}. Each simulation uses the same initial value for $\alpha$, $1.75$. They use different collision outcome prescriptions, noted in the plot titles. The upper panel of each plot shows the stellar density as a function of distance from the SMBH at a particular time. As indicated by the colorbar on the right, the redder curves correspond to older populations. The bottom panel shows the fractional change in stellar density due to collisions compared to the expected value. The black line shows the original, reference density profile. 

In each case, collisions deplete the stellar mass near the SMBH. This process causes the density profile to flatten within $r_{\rm coll}$. Over time, this $r_{\rm coll}$ moves further from the SMBH, shifting the break radius in the density profile. The fully destructive case in the left column of Figure~\ref{fig:densityprofiles} provides the clearest example of the break radius sweeping outward over time: the bluest density profiles diverge from the unmodified profile at smaller radii. Equating timescales suggests a break radius of $\sim 0.1$~pc for a $\sim 10$ Gyr population with these initial conditions (Eq.~\ref{eq:scale2}). The break radius of the fully destructive model falls slightly outside of this radius, closer to $0.3$~pc for a $7$ Gyr old population, because collisions still affect a fraction of the population outside of $r_{\rm coll}$ leading to mass loss for some percentage of the stars (see Figure 1 in \citet{Rose+23}), causing break radii to be gradual rollovers rather than abrupt transitions. 

A more realistic mass loss and merger prescription gives a less distinct break radius, as seen in the second two columns of Figure~\ref{fig:densityprofiles}. In these prescriptions, a single collision results in comparatively smaller fractional changes in total stellar mass. Compared to the Rauch99 model, the Lai+93 density curves show greater flattening within $0.05$~pc. The greater flattening occurs because fitting formulae from the corresponding study \citet{Lai+93} give a higher fractional mass loss per collision in this region compared to \citet{Rauch99}. Mergers can cause also contribute to a break in the density profile. While mergers result in fractional mass loss, generally between five to ten percent \citep[see figure 2 in][]{Rose+23}, they also hasten the evolution of the stars off the main-sequence. As a result, there is less mass in main-sequence stars at late times within $0.1$~pc compared to the outer region of the nuclear star cluster.

We also examine the dependence of the break radius on the initial density profile assumed for the population. We test three values of $\alpha$: $1.25$, $1.5$, and $1.75$.  Figure~\ref{fig:differentalpha} juxtaposes the evolving density profiles for three simulations with these different values of $\alpha$. These simulations use the same collision outcome prescription, Rauch99. We find that a break in the density profile is ubiquitous, but its location depends on $\alpha$, shifting from $\sim 0.1$~pc for $\alpha =1.75$ to $\sim 0.04$~pc for $\alpha = 1.25$.  The steeper the initial profile, the greater the fractional change in density, as seen in the bottom panels of the Figure. To illustrate these trends, we indicate the point at which the density at $7$ Gyr is halved compared to the expected value using a grey vertical line in each panel. It shifts inward with increasing $\alpha$. Regardless of initial conditions, we find that the density profile is always preserved, i.e. unmodified by collisions, outside about $0.2$~pc. Observations of this region can therefore be used to constrain the underlying, original density distribution.

\subsection{Cluster Luminosity and Color} \label{sec:luminosity}

As we discuss in the previous section, mass loss due to collisions together with the accelerated evolution of merger products can lead to a discernible flattening of the stellar density profile over time. In this section, we briefly address the effect of collisions on the stellar luminosity. Mergers are the typical outcome of collisions outside of about $0.01$~pc \citep{Rose+23}. Consistent with our previous treatment of the merger products, we assume that the properties of the merged star are similar to those of a typical main-sequence star of the same mass \citep[e.g.][]{2019ApJ...881...47L}. In this case, the luminosity of the stars should follow a simple mass-luminosity relation: $L \propto M^{2.3}$ for $M<0.43 \, M_\odot$, $\propto M^4$ for Sun-like stars, and $\propto M^{3.5}$ for $M>2 \, M_\odot$ \citep[e.g.,][]{SalarisCassisi,Duric03}. Additionally, we can calculate the peak wavelength for the stars in our sample with Wien's displacement law, which we use the peak wavelength as a proxy for color. The general assumption here is that we consider the properties of these objects after a period of thermal relaxation, not when they are still cooling following collisional shock-heating \citep[see discussion of thermal versus collision timescale in][]{Rose+23}.

\begin{figure}
	\includegraphics[width=0.5\textwidth]{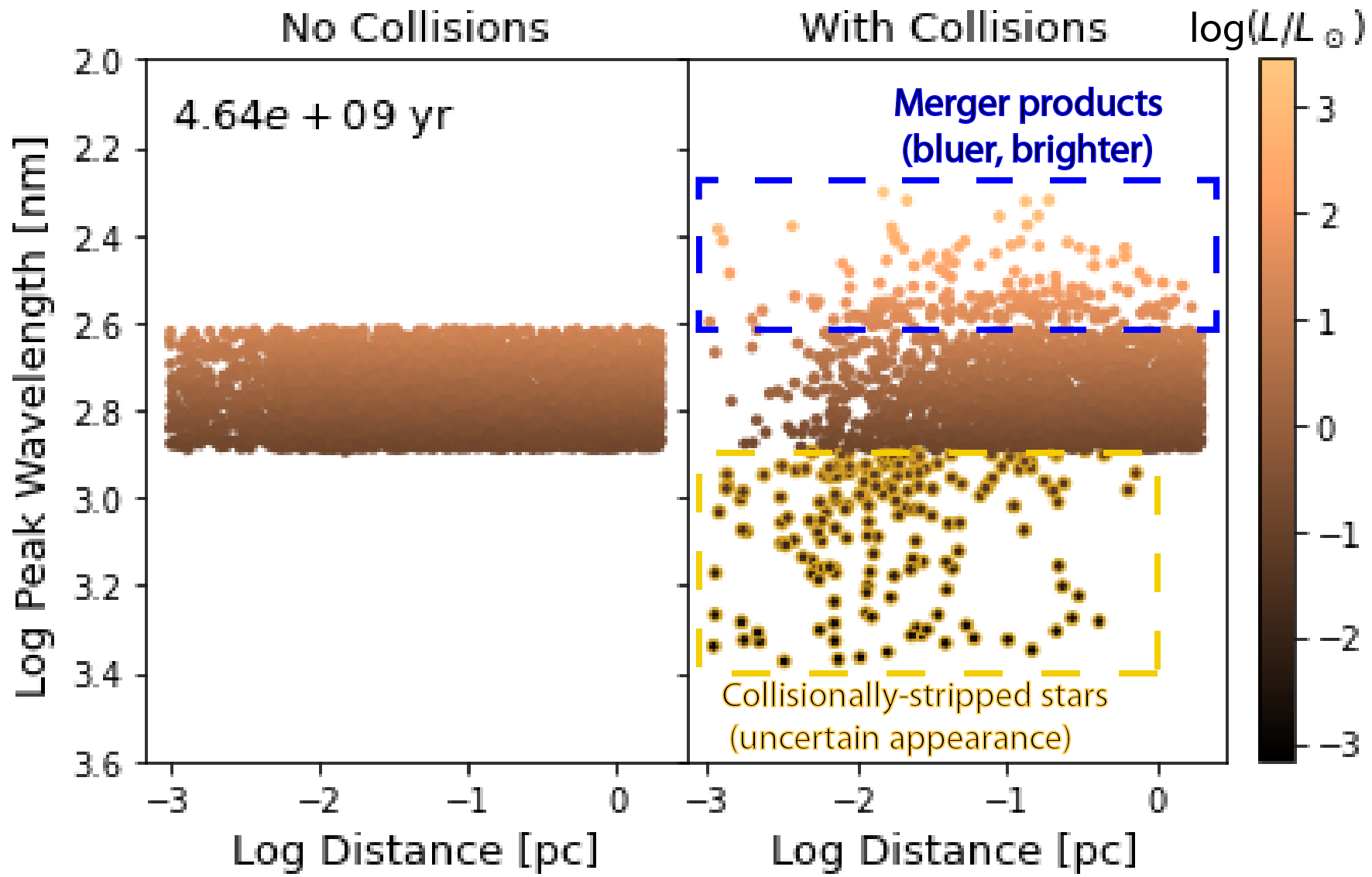}
	\caption{The peak wavelength of each star in our sample $4.64 \times 10^9$~yr into our Rauch99, $\alpha = 1.75$ simulation. Peak wavelength, a proxy for color, is plotted versus distance from the SMBH. We also colorcode the stars in the sample by bolometric luminosity. The left panel shows the stars without collisions, the same as the grey points in the lower right plot of Figure~\ref{fig:snaps}, while the right panel shows the simulated population with collisions. Collision-induced mergers cause a population of brighter, bluer stars to form, stars which would not otherwise exist in a population of the same age. We also include the population of collisionally-stripped stars, though the luminosity, color, and general appearance of these stars is uncertain.
 }
     \label{fig:luminosity}
\end{figure}

We illustrate the effects of collisions on color and luminosity in Figure~\ref{fig:luminosity}. This figure presents a snapshot of our sample stars at $4.64 \times 10^9$~yr. The left panel shows the expected population in the absence of collisions, equivalent to the grey points in Figure~\ref{fig:snaps}, while the right side shows the simulated population. We plot the peak wavelength of the stars versus their distance from the SMBH, both on a logarithmic scale. The y-axis is inverted so that more massive (bluer) stars are higher, facilitating a direct comparison with the mass versus radius plot in Figure~\ref{fig:snaps}. Additionally, we color-code the points by bolometric luminosity, calculated using the relations above. 

As can be seen in the figure, mergers  produce brighter and bluer stars than expected for a population of that age, like ``blue stragglers" in a star cluster population \citep{2019ApJ...881...47L}.
Collisionally stripped stars are also present in Figure \ref{fig:luminosity}.  These stars have undergone one or more high-speed collisions, leading to mass loss. Their new, lower masses would suggest that they are redder and less luminous. However, while we have plotted them here based on this assumption, their appearance is highly uncertain. Models of stripped stars in binary systems may provide clues as to their appearance, suggesting they are in fact more luminous than their progenitor stars \citep{Gotberg+18}.  We therefore distinguish them with gold outlines in the figure.

Previously, we showed that collisions always produce a break in the stellar mass density profile. The break occurs because collisions can only ever reduce the mass contained in the stars; it cannot increase the mass. Luminosity, however, scales as mass to the $3.5$ power. We therefore do not expect collisions to necessarily produce a decrease in the bolometric luminosity profile of the cluster within $r_\mathrm{coll}$. The bluer, brighter merger products may in fact outshine the other stars in their vicinity. Figure \ref{fig:luminosity} shows that the colors and luminosities of stars are affected by collisions in a way that varies systematically with radius. Luminous, blue merger remnants exist from 0.01--1~pc, around  $r_{\rm coll} \sim 0.1$~pc.  Stripped stars from the highest-velocity collisions preferentially lie within $0.1$~pc. Future work may examine the evolution of the luminosity profile, both bolometric and in specific bands, by leveraging comprehensive hydrodynamic and stellar evolution simulations in order to understand the dynamical and thermal evolution of merger products. 


\section{Conclusions} \label{sec:conclusions}

Collisions between main-sequence stars are common within a tenth of a parsec of the SMBH in the center of the Milky Way galaxy. The impact velocities of these collisions are often on the order of, if not larger, than the escape speed from the stars \citep[e.g.,][]{Lai+93,Balberg+13,Rose+23}. As a result, individual collisions can result in mass loss from the stars \citep[e.g.,][]{Lai+93}. On a population level, mass loss from collisions can affect the stellar density profile of the cluster, or feed the central SMBH \citep[e.g.][]{2011AdAst2011E..13R}. Some key findings of our work include:
\begin{enumerate}
    \item Collisions affect the majority of stars inside the collision radius, $r\lesssim r_{\rm coll}$, defined by $t_{\rm coll} =  t_{\rm age}$. In our GN, $r_{\rm coll}\sim 0.1$~pc (equation \ref{eq:scale2}).   
    \item As described in further detail by \citet{Rose+23}, lower-velocity collisions lead to massive, ``blue straggler" merger remnants, while the highest-velocity collisions lead to stripped, low-mass remnants. The occurrence of these products depends on radius within the cluster, with the majority at $r\lesssim r_{\rm coll}$ (Figures \ref{fig:snaps} and \ref{fig:luminosity}).
    \item Collisions always result in at least partial stellar mass loss. Additionally, many collisions can merge stars into more massive stars between $0.01$ and $0.1$~pc. These stars evolve off the main-sequence more quickly, creating a deficit in stellar mass compared to regions further from the SMBH. We examine the effect of collisions on the stellar mass density profile in Figures \ref{fig:densityprofiles} and \ref{fig:differentalpha}. Inside a break radius, $\sim r_{\rm coll}$, density profiles decrease from their nominal values. We find that the location slope of the stellar mass density inside the break radius depends most-strongly on the collision model adopted and how much mass is expelled in high velocity collisions. 
\end{enumerate}
Our results demonstrate a simple, intuitive relation between the location of $r_{\rm coll}$ and the density profile and individual properties of stars in GN. Equation~\ref{eq:scale_simple} highlights how these results can be extrapolated to other systems. Our findings highlight how future work could address key uncertainties by exploring the interplay of dynamics, hydrodynamics, and stellar evolution in this unique astrophysical setting.

\begin{acknowledgments}
We thank Smadar Naoz and Abraham Loeb for many helpful discussions. SR thanks the Dissertation Year Fellowship, Bhaumik Institute Fellowship, and CIERA Lindheimer Fellowship for partial support. SR acknowledges partial support from NASA ATP 80NSSC20K0505 and NSF-AST 2206428 grants. MM is grateful for support from a Clay Postdoctoral Fellowship at the Smithsonian Astrophysical Observatory. 

\end{acknowledgments}






\bibliography{sample631}{}
\bibliographystyle{aasjournal}



\end{document}